\documentclass[oneside,12pt]{amsart}
\usepackage{amssymb,amscd,amsmath,latexsym}
\usepackage{amsthm}
\usepackage{layout}
\usepackage[mathcal]{euscript}
\usepackage{graphicx}
\usepackage[all]{xy}
\CompileMatrices

\def\U{{\bf U}}

\begin{document}

\title {Game Theory in Categorical Quantum Mechanics}

\author[Ali Nabi Duman]{Ali Nabi Duman}
\address{Department of Mathematics,
College of Science, University of Bahrain, Bahrain} \email{aduman@uob.edu.bh}

\date{\today}

\maketitle
\def\SS{{\mathbb S}}
\def\G{{\mathcal G}}
\def\B{{\mathcal B}}
\def\H{{\mathcal H}}
\def\N{{\mathcal N}}
\def\K{{\mathcal K}}
\def \x{{\bf x}}
\def \M{{\mathcal M}}
\def \C{{\mathbb C}}
\def\HH{{\mathbb H}}
\def \Z{{\mathbb Z}}
\def \R{{\mathbb R}}
\def \Q{{\mathbb Q}}
\def \U{{\mathcal U}}
\def \E{{\mathcal E}}
\def \z{{\bf z}}
\def \m{{\bf m}}
\def \k{{\bf k}}
\def \n{{\bf n}}
\def \g{{\bf g}}
\def \h{{\bf h}}
\def \V{{\mathcal V}}
\def \W{{\mathcal W}}
\def \T{{\mathbb T}}
\def \X{{\mathcal X}}
\def \Y{{\mathcal Y}}
\def \P{{\bf P}}
\def \F{{\bf F}}
\def \p{{\mathfrak p}}
\def \LL{{\mathfrak L}}
\def \L{{\mathcal L}}
\def \O{{\mathfrak O}}
\def \longto{\longrightarrow}
\def \sl{{\frak sl}}
\def \C{{\mathbb C}}
\def \Q{{\mathbb Q}}
\def \vCech{{\v{C}ech\ }}
\def \calP{{\mathcal P}}
\def \H{{\mathcal H}}
\def \Cbar{{\bar{\mathcal C}}}
\def \N{{\mathbb N}}
\def \S{{\Sigma}}
\def \s{{\sigma}}
\def \T{{\mathbb T}}
\def \t{{\tau}}
\def \z{{\bar z}}
\def \bZ{{\bar Z}}
\def \P{{\mathbb P}}
\def \R{{\mathbb R}}
\def \HH{{\mathbb {H}}}
\def \div{{\rm div}}
\def \W{{\mathbb W}}
\def \L{{\mathcal L}}
\def \l{{\lambda}} \def \bl{{\Lambda}}
\def \bu{\bullet}
\def \fd{{\bullet}}
\def \Z{{\mathbb Z}}
\def \e{{\varepsilon}}
\def \ag{{\frak{g}}}
\def \ah{{\frak{h}}}
\def \O{{\mathcal O}}
\def \M{{\mathcal M}}
\def \K{{\mathcal K }}
\def \CC{{\mathcal C}}
\def \XC{{{X}_{\mathcal C}}}
\def \MC{{{\mathcal M}_{\mathcal C}}}
\def \cf {{\mathcal F}}
\def \w{{\wedge}}
\def \x{{\bf x}}
\def \k{{\kappa}}
\def \XCbar{{{X}_{\bar{\mathcal C}}}}
\def \MCbar{{{\mathcal M}_{\bar{\mathcal C}}}}
\def \i{{\sqrt{-1}}}
\def \p{{\partial}}
\def \b{{\delta}}
\def \D{{\Delta}}
\def \G{{\mathcal G}}
\def \o{{\omega}}
\def \g{{\gamma}}
\def \re {\noindent {\it Remark\ \ }}
\def \proof{{\noindent{\it Proof.\ \ }}}

\newtheorem{Th}{Theorem}[section]
\newtheorem{cor}[Th]{Corollary}
\newtheorem{lem}[Th]{Lemma}
\newtheorem{prop}[Th]{Proposition}
\newtheorem{claim}[Th]{Claim}

\theoremstyle{definition}
\newtheorem{dfn}[Th]{Definition}
\newtheorem{example}[Th]{Example}
\theoremstyle{remark}
\newtheorem*{remark}{Remark}
\newtheorem{lemma}[Th]{Lemma}

\begin{abstract}

Categorical quantum mechanics, which examines quantum theory via dagger-compact closed categories, gives satisfying high-level explanations to the quantum
information procedures such as Bell-type entanglement or complementary observables (\cite{AC}, \cite{Co}, \cite{Co2}). Inspired by the fact that Quantum
Game Theory can be seen as branch of quantum information, we express Quantum Game Theory procedures using the topological semantics provided by Categorical
Quantum Mechanics. We also investigate Bayesian Games with correlation from this novel point of view while considering the connection between Bayesian game
theory and Bell non-locality investigated recently by Brunner and Linden \cite{BL}.

\end{abstract}

\section{Introduction}

The standard axiomatic presentation of quantum mechanics in terms of Hilbert spaces was established by von Neumann about 80 years ago \cite{vN}. From this
point of view, a quantum procedure can be described by state preparations, unitary operators and projective measurements utilizing matrices of complex
numbers. As emphasized by Vicary \cite{V}, the aim of this description is to implement a protocol rather than providing an insightful explanation about its
mechanism.

With the introduction of the field \emph{quantum information and computation}, a need to answer new type of questions and to revisit the foundation of the
quantum mechanics is arisen. Hence the standard methods of quantum mechanics turns out to be deficient for developing quantum algorithms and protocols.
According to Abramsky and Coecke \cite{AC2}, there are two main disadvantages related to the standard tools in use: Firstly, they are too low-level to
introduce the modern Computer Science concepts such as types, abstraction and the use of the tools from algebra and logic. Secondly, they are not
comprehensive enough to describe quantum protocols such as teleportation where the outcome of the measurement is the main way to determine the actions of
the process. Abramsky and Coecke pioneered Categorical Quantum Mechanics (CQM) programme in order to address these problems. Their main mathematical
setting is based on symmetric monodial categories which is non-surprisingly the same mathematical structure used by Lambek \cite{L} to describe the
interfaces and implementations in object-oriented programming.

Another field that has been adopting ideas from quantum information and computation is game theory. Game theory is the study of decision making in conflict
situations. It has been widely used in social sciences, economics and biology. Modern Game Theory was first introduced by von Neumann and Morgenstern
\cite{VM} in 1944 and mainly formalized by the work of John Nash in the following years. In \cite{VM}, von Neumann and Morgenstern elaborate the ideas from
physics to clarify the economical concepts while defying the premature objection of social scientists stating that an economic theory cannot be modelled
after physics.

Recent developments in quantum information gave birth to the field Quantum Game theory. Quantum game theory is the study of strategic behavior of agents
with the access to quantum technologies such as entanglement, teleportation etc.. There are two ways to utilize quantum technologies in a game: They can be
used for randomization of the game or as a communication protocol between the agents \cite{La}.

In the case of randomization, the players coordinate their strategies via quantum devices. The corresponding equilibria form a subclass of correlated
equilibria in the sense of Aumann \cite{A}. We can consider the recent work of Brunner and Linden \cite{BL} from this perspective. In their work, they
discuss the connection between Bell nonlocality and Bayesian Games while formulating a Bayesian Game as Bell inequality test scenario where the agents use
a common advise allowing for correlated strategies. Using nonlocal resources, such as entangled particles, the players can achieve better equilibria called
quantum Nash equilibria or non-signaling Nash equilibria. These equilibria are equivalent to violating the Bell inequalities in the quantum mechanics
setting.

Alternatively, the quantum technologies can be used as a mean for communication between the agents. This results in a new set equilibria which has no
classical interpretation in game theory because the communication protocol is usually not specified in a game. However, Meyer \cite{M} pointed out that
quantum communication technologies can effect the outcome of the game. He also showed that this outcome of the game can change according to the type of the
communication protocol. One of the most studied protocols is the Eisert-Wilkens-Lewenstein \cite{EWL} protocol which illustrates effect of quantum
communication in a general context.

The aim of this paper is to introduce Categorical Quantum Mechanics to the field of Quantum game theory. We address EWL-protocol as well as correlated
Bayesian Games from this new point of view. This enable us to revisit the both aspect from Quantum game theory mentioned above.

The paper is organized as follows: In section 2, after a brief introduction of EWL-model, we use categorical diagrammatic to present it. In section 3, we
focus on correlated Bayesian Games in CQM emphasizing the connection with Bell non-locality. Finally, section 5 draws the conclusion.

\section{Quantum Communication in Categorical Quantum Mechanics}

In a game, the player should communicate their strategies to calculate their payoffs at the end of the game. In most of the cases, this communication
protocol is not modelled. One can achieve the communication through a referee. In this case, a referee hands the players pennies which the players can
transform from one state to another.At the end, the pennies are returned to the referee who computes payoffs. In the real life cases the referee can be
considered as a marketplace or an arbiter. In this section, we consider  the case where the communication is achieved by using quantum technologies.

\subsection{Eisert-Wilkens-Lewenstein Protocol}

EWL-protocol is one of the most referred protocols for quantum games as it demonstrates the effects of quantum communication in a general context. We first
explain the setting of quantum game for this protocol. One can specify a quantum game $\Gamma=(\mathcal{H}, \rho, S_A, S_B, P_A, P_B)$ by the Hilbert space
$\mathcal{H}$, the initial state $\rho$, the strategy spaces $S_A$ and $S_B$ and payoff functions $P_A$ and $P_B$ for each player $A$ and $B$. The quantum
strategies $s_A \in S_A$ and $s_B \in S_B$ are unitary operators mapping the state space on itself. The object of the game is to determine strategies
maximizing the payoffs to a player. The initial state $\rho$ is known to both players $A$ and $B$. After choosing their strategies $s_A$ and $s_B$ the
final state $$\sigma=(s_A\otimes s_B)\rho$$ is computed. Next, the referee perform the projective measurement on the final state $\sigma$ and compute the
payoffs for each player. One can present EWL-protocol for two qubits in circuit diagram as follows:

\begin{center}
\includegraphics[scale=1]{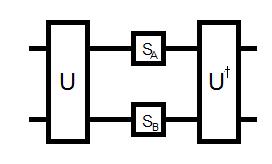}
\end{center}

In this case we apply unitary operator $U$ to obtain an entangled state $\rho=U |00\rangle$ and inverse of $U$ is applied at the end to bring the game
 to the final state $\sigma=U^{\dagger}(s_A\otimes s_B) U|00\rangle$.

 We now briefly explain the quantum version of Prisoners' Dilemma \cite{EWL}. This procedure can be applied to any strategic-form game. We map corporation and defect vectors $|C\rangle$ and $|D\rangle$ to $|0\rangle$ and $|1\rangle$,
 respectively. The measurement of the system is projected into one of the four vectors $|00\rangle$, $|01\rangle$, $|10\rangle$ and $|11\rangle$, with
 associated probability, resulting in the following payoff $P_A$ for player $A$:
$$P_A=3|\langle \sigma |00\rangle|^2 +0|\langle \sigma |01\rangle|^2+5|\langle \sigma |10\rangle|^2+1|\langle \sigma |11\rangle|^2.$$
In order to obtain an entangled state, we choose our unitary operator to be
$$U=\frac{1}{\sqrt{2}}(I^{\otimes 2}+i\sigma_x^{\otimes 2})$$ where $\sigma_x$ is the Pauli operator.
Its inverse is
$$U^{\dagger}=\frac{1}{\sqrt{2}}(I^{\otimes 2}-i\sigma_x^{\otimes 2}.)$$

If the strategy spaces $S_A$ and $S_B$ consist of the identity matrix $I$ and Pauli operator $\sigma_x $ we obtain the same payoff matrix as classical
Prisoners' Dilemma game. However, if we allow Hadamard operator move then we get different outcomes. For example, for $s_A \sim I$ and $s_B \sim H$ the
final stage is $$\sigma=U^{\dagger}(I\otimes H) U|00\rangle =\frac{1}{\sqrt{2}}(|01\rangle - i|11\rangle).$$Hence the payoffs in this case is $P_A=0.5$ and
$P_B=3$. One can easily verify that the Nash equilibrum corresponding to this game is $(H,H)$ even though it is still not Pareto optimal. On the other
hand, introducing the Pauli matrix $\sigma_z$ to the strategy spaces allows us to obtain the pareto optimal Nash equilibrum $(\sigma_z, \sigma_z)$.
\subsection{Topological Semantics}

In this subsection, we give the necessary background for topological semantics to present the EWL-protocol. This new perspective has already provide
fruitful results in giving high-level description to quantum procedures. Assuming quantum game theory as a branch of quantum information, we hope that the
topological presentation of quantum games enable us to give more insight in explaining the procedures of the games as it does in the case of quantum
algorithms \cite{V}.

The content of this section can be found in the appendix of \cite{V}. This diagrammatic is widely used in quantum foundation and information \cite{Co}. One
can also refer to \cite{KL} and \cite{S} for the mathematical foundation of the notation given by category theory.

We start with the identity map on a finite Hilbert spaces. This is represented by a vertical wire.
\begin{center}
\includegraphics[scale=0.3]{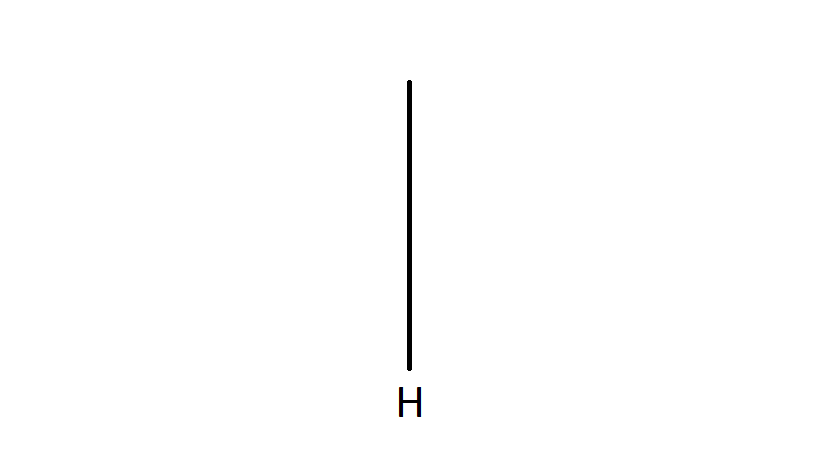}
\end{center}

The following diagram represent a linear map $p:H \rightarrow J.$
\begin{center}
\includegraphics[scale=0.3]{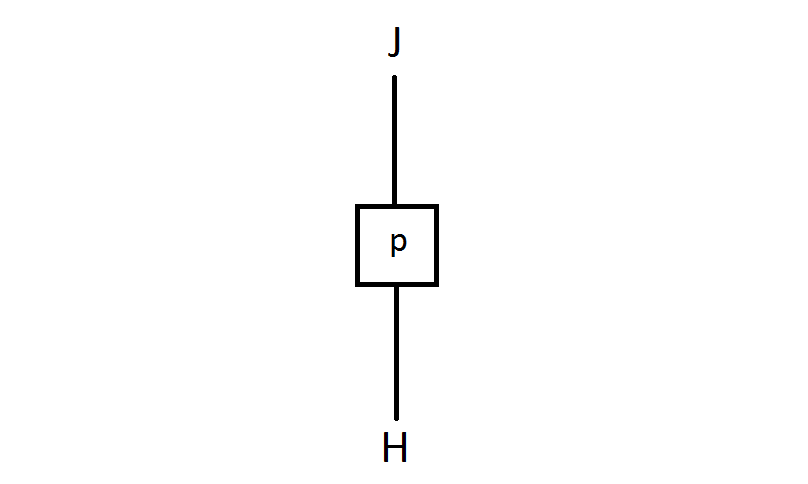}
\end{center}
 Horizontal juxtaposition of diagrams represents tensor product of linear maps, and vertical juxtaposition represents composition of linear maps.

The identity on the $1$-dimensional Hilbert space is represented as the empty diagram: \vspace{2cm} \newline One can change the relative heights of the
boxes and move the components around.
\begin{center}
\includegraphics[scale=0.5]{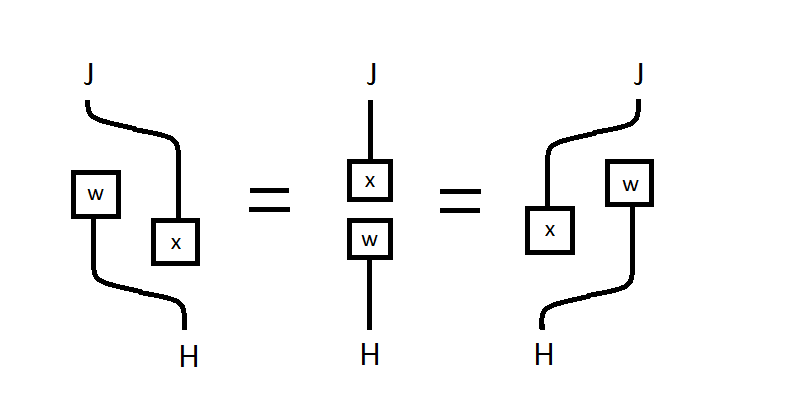}
\end{center}

A chosen vector $x\in H$ corresponds to a map $x:1\rightarrow H$, which we denote graphically as: \begin{center}
\includegraphics[scale=0.5]{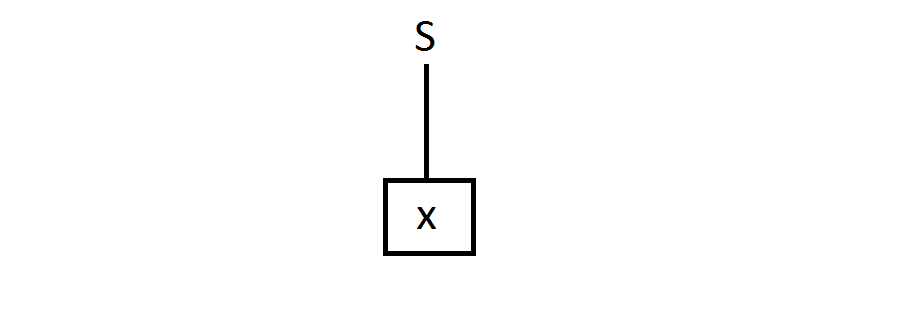}
\end{center} This box without a bottom line corresponds to the preparation while a box without top line corresponds to the measurement.
\subsection{EWL-protocol}

 Like in the case of recent quantum procedures, quantum circuit model of quantum computation might not be comprehensive enough to describe the quantum games. One can
 raise following question regarding the EWL given above:

\begin{itemize}
\item How does the strategy space change the outcome of the game?

\item Does the initial state $|00\rangle$ and unitary operator $U$ have an important role?

\item Are there any other equivalent presentations of the game?
\end{itemize}

In order to look these questions from a different point of view, we now present the topological structure of EWL-protocol making use of the topological
formalism for linear algebra given in the last subsection.
\begin{center}
\includegraphics[scale=0.7]{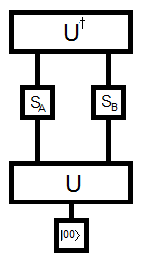}
\end{center}

The operators $U$ and $U^{\dagger}$ can be considered as a part of preparation and the measurement, respectively. $s$ is the projective selective
measurement. Here we can decompose $U$ as linear combination of two diagrams as follows:

\begin{center}
\includegraphics[scale=0.7]{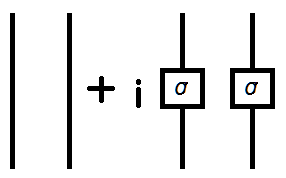}
\end{center}
We compose this diagram with $s_A\sim I$ and $s_B \sim H$ in the following way:
\begin{center}
\includegraphics[scale=0.7]{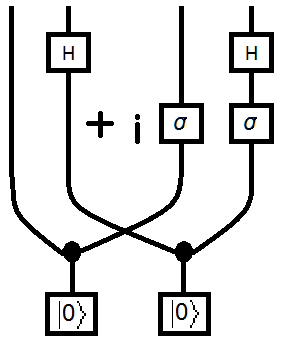}
\end{center}
 Here each players' move has an effect on the entangled state rather than on their own qubit separately and the function of $U^{\dagger}$ is to
allow effective projective measurement. Indeed, one can also consider mixed quantum state instead of probability distributions as the outcome of the game.

One present $N$-player game as follows:
\begin{center}
\includegraphics[scale=0.7]{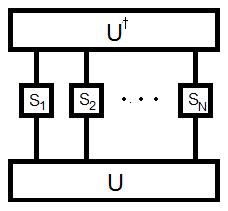}
\end{center} In this case, one can use any maximally entangled initial state (MEIS) such as W-state or GHZ-state. Chapell et. al. \cite{CIA} noticed that
in the case Prisoners' Dilemma game the outcomes are the same for GHZ-state and W-state when number of players is equal to $2$. It is known that these two
states are equivalent under local operators.

In order to investigate the effect of the initial state in CQM setting, one can refer to the work of Coecke and Kissinger \cite{CK} where they expose the
graphical and algebraic structure of the GHZ-state and the W-state, as well as a purely graphical distinction that characterizes the behavior of these
states.

\section{Quantum Randomization in Categorical Quantum Mechanics}

In order to model certain real life situations, a set of $S_i$-valued random variables $\mathcal{X}_i$ is assigned to each player. These variables are not
necessarily independent. The elements of $\mathcal{X}_i$ are called \emph{types} and $(\mathcal{X}_1,\ldots, \mathcal{X}_n)$ is called the
\emph{environment}. The pair of random variables $(X_1,X_2)$ is called a \emph{correlated equilibrum} if it is Nash equilibrum in the game
$G(\mathcal{X}_1, \mathcal{X}_2)$. Two correlated equilibra are \emph{equivalent} if they induce the same probability distribution on the strategy spaces
$S_1\times S_2$.

The theory of quantum strategies models the behavior of players with access to quantum randomizing devices. Here we replace the sets $\mathcal{X}_i$ of
random variable with the sets $\mathcal{X}_i$ of quantum mechanical observables. As with correlated equilibra, if $(X_1, X_2)$ is a Nash equilibrium in the
game $G$, $(X_1, X_2 )$ is called a \emph{quantum equilibrium} in $G$. In a similar manner, two quantum equilibria are called \emph{equivalent} if they
induce the same probability distribution on the strategy spaces $S_1\times S_2$. Moreover, If $(X, Y )$ is a quantum equilibrium then (by the definition of
quantum environment), X and Y are simultaneously observable, and hence can be seen as a classical random variable. In the next subsection we summarize the
results of Linden and Brenner stating the connection between Bell non-locality and Bayesian Games \cite{BL}.

\subsection{Bell non-locality and Bayesian Game Theory}
Non-locality, which is one of the most counter-intuitive  features of quantum theory, states that two remote observers sharing a pair of entangled
particles can establish correlations which is beyond the explanation of classical physics. This phenomenon, also confirmed experimentally via violation of
Bell inequalities, is proved to be useful in practical areas such as quantum information.

On the other hand, Bayesian games formulated by Harsanyi are the games in which players have partial information about the setting of the games. Bayesian
game theory plays an important role in economics used particularly to model auctions.

In \cite{BL}, Brenner and Linden discuss the connection between Bell non-locality and Bayesian games by reformulating the normal form of a Bayesian game as
a Bell inequality test scenario. In this type of setting the players receive advice in the form of non-local correlations such as entangled particles or
non-signaling boxes. This quantum resources offer better outcome than the classical ones. This advantage is first discussed by Cheon and Iqbal \cite{CI}
where payoff function corresponds to Bell inequality. In the case under consideration, none of the payoff functions corresponds to a Bell inequality. This
is in contrast with the approaches discussed in the previous section where quantum advantage is achieved only under specific restrictions.

The normal form representation of a Bayesian game is given by the following ingredients:
\begin{itemize}
\item The number of players $N$.

\item A set of states of nature $\Omega$, with a prior $\mu(\Omega)$

\item For each player $i$, a set of strategies $S_i$.

\item For each player $i$, a set of types $\mathcal{X}_i$.

\item For each player $i$, a mapping $\tau_i:\Omega \rightarrow \mathcal{X}_i$

\item For each player $i$, a payoff function $P_i:\Omega \times S_1 \times \ldots \times S_N \rightarrow \mathbb{R}$, determining the score of the player
for any possible combination of types and actions.
\end{itemize}
The average payoff for each player $i$ is given by
$$F_i=\sum \mu(X_1,\ldots , X_N)p(s_1, \ldots, s_N|X_1,\ldots , X_N)P_i(X_1,\ldots , X_N, s_1, \ldots, s_N)$$
where the sum goes over all variables $X_1,\ldots , X_N, s_1, \dots, s_N$. $p(s_1, \ldots, s_N|X_1,\ldots , X_N)$ is the probability of the strategies
$s_1, \ldots, s_N$ of given type $X_1,\ldots , X_N$. In the case of correlated classical advice, the advice is represented by a classical variable,
$\lambda$, with the prior $\rho(\lambda)$. We have
$$p(s_1, \ldots, s_N|X_1,\ldots , X_N)=\sum_{\lambda} \rho(\lambda)p(s_1|X_1, \lambda)\ldots p(s_N|X_N, \lambda)$$

One can analyze a game using the set of payoff functions $\{F_1, \ldots, \F_N\}$ considering all possible strategies. In the case of classical advice the
set of points in $\mathbb{R}^n$ with coordinates $(F_1, \ldots, \F_N)$ is a convex polytope characterized with the following inequalities
$$\sum_{i=1}^{N} \beta_i \F_i \leq \beta_0$$
where $\beta_i$ are real numbers.

The above setting is related to the Bell test scenario. One can formulate this for $N$ parties in the following way: There are $N$ parties $A_i$ sharing a
physical resource distributed by a central source. Each observer receives a question(measurement), $X_i$, which he is asked to give an answer(outcome of a
measurement), $s_i$. In the setting of a Bayesian game the questions and answers correspond to types and strategies, respectively. After repeating this
experiment large number of times the statistics of a game can be computed by the probability distribution
$$p(s_1, \ldots, s_N|X_1,\ldots , X_N)$$
which is the probability of observing the answers $s_1, \ldots, s_N$, given the questions $X_1,\ldots , X_N$. In the case of classical source the
statistics can be written as
$$P(s_1, \ldots, s_N|X_1,\ldots , X_N)=\int d\rho(\lambda) \rho{\lambda}p(s_1|X_1, \lambda)\ldots p(s_N|X_N, \lambda)$$
where the variable $\lambda$ is the information distributed from the source to all observers. In a Bayesian game, $\lambda$ corresponds to an advice. Bell
discovered that the correlations in an experiment involving a classical source is constraint:
$$\sum_{s_1, \ldots, s_N, X_1,\ldots , X_N}\alpha_{s_1, \ldots, s_N, X_1,\ldots , X_N}p(s_1, \dots, s_N|X_1,\ldots , X_N)\leq L$$
where $\alpha_{s_1, \ldots, s_N, X_1,\ldots , X_N}$ are real numbers. Brenner and Linden noticed that the payoff function has the same form with
$$\alpha_{s_1, \ldots, s_N, X_1,\ldots , X_N}=\mu(X_1,\ldots , X_N)P_i(X_1,\ldots , X_N, s_1, \ldots, s_N).$$

In the existence of quantum particles as central source, the Bell inequality will be violated. For Bayesian games, this corresponds to the fact that if the
players have access to non-local advice they outperform any classical players as the statistics of the non-local measurement cannot be reproduced by any
classical model. In other words, a Bayesian game with classical advice is not equivalent to a Bayesian game with quantum advice. In the last subsection, we
show this fact via categorical quantum mechanics.

\subsection{Categorical quantum mechanics revisited}

In \cite{Me}, Mermin establishes non-locality as a contradiction of parities rather than as a violation of Bell inequality. Using Mermin argument, Coecke
and et. al. \cite{Co3} provide new insight as well as generalization of non-locality in the context of categorical quantum mechanics. We now give the
necessary language that we will use to present a quantum game in the sense of the previous subsection. One can address \cite{Co3} for more detailed
discussion.

An observable yields classical data from a physical system. In quantum mechanics an observable in a self-adjoint operator. The information encoded by an
observable is eigenvectors. In the category of finite dimensional Hilbert spaces, the orthonormal basis is 1-to-1 corresponce  with $\dagger$-special
commutative Frobenius algebra. In a $\dagger$-symmetric category(SMC) a $\dagger$-special commutative Frobenius algebra ($\dagger$-SCFA) is a commutative
Frobenius algebra
$$\mathcal{O}_\circ=(\mu_\circ: X\otimes X \rightarrow X, \eta_\circ: I\rightarrow X, \delta:X\rightarrow X\otimes X, \epsilon_\circ : X\rightarrow I)$$
such that $\delta_\circ = (\mu_\circ)^{\dagger}$, $\epsilon_\circ = (\mu_\circ)^{\dagger}$. We can denote $\mu_\circ$, $\eta_\circ$, $\delta_\circ$ and
$\epsilon_\circ$ pictorially as follows:

\begin{center}
\includegraphics[scale=0.7]{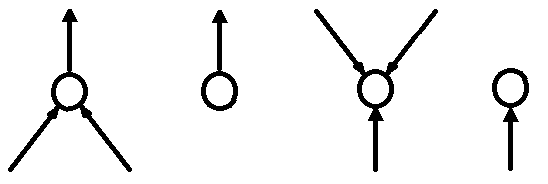}
\end{center} Each observable structure comes with a set of classical points, the abstract analogues to eigenvectors of an observable:

\begin{center}
\includegraphics[scale=0.7]{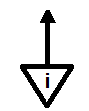}
\end{center} A multiplication puts a monoid structure on the points of $X$. If restrict those points $\psi_\alpha : I\rightarrow A$ we obtain an abelian
group $\phi_\circ $ called the phase group $\mathcal{O}_\circ$. We represent these points as:

\begin{center}
\includegraphics[scale=0.7]{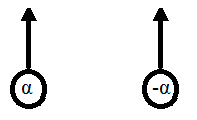}
\end{center}
A measurement is defined as:
\begin{center}
\includegraphics[scale=0.7]{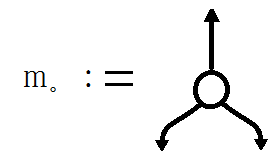}
\end{center}
 A point $|\Gamma): I\rightarrow X$ of the following form is called \emph{Born vector}

\begin{center}
\includegraphics[scale=0.7]{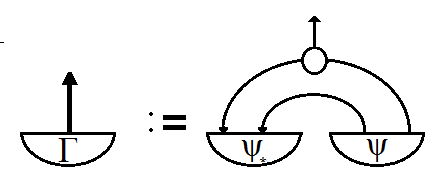}
\end{center}where
\begin{center}
\includegraphics[scale=0.7]{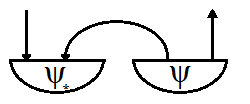}
\end{center} is a quantum state represented by a positive operator using Selinger's representation for completely positive maps . One can extend the above
definition to the points of the form $|\Gamma): I\rightarrow X\otimes \ldots \otimes X$.

\subsection{Bayesian games in CQM}

The classical probability distribution for $N$-measurements against arbitrary phases $\alpha_i$ on $N$ systems of any type of generalized
$GHZ^N_\circ$-state is given as follows:

\begin{center}
\includegraphics[scale=0.7]{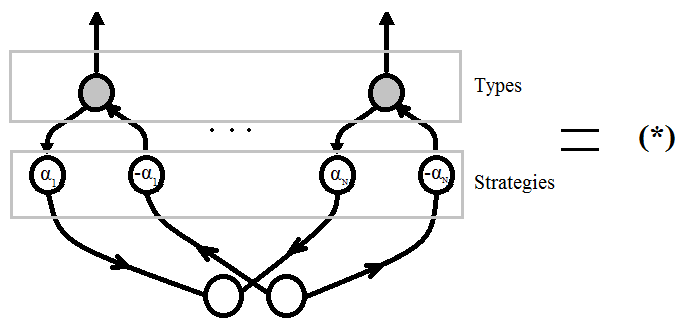}
\end{center} Using the analogy made by Linden and Brenner \cite{BL}, we can conclude the orthonormal basis, where the projective measurements are done,
correspond to the types and the $N$-tuple $(\alpha_1,\ldots,\alpha_N)$ correspond to the strategies in a Bayesian game with $N$ players. This distribution
will be equal to \cite{Co3}:

\begin{center}
\includegraphics[scale=0.7]{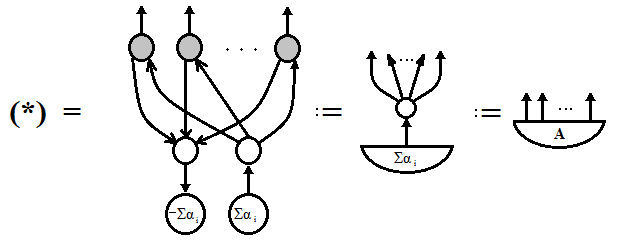}
\end{center} In other words, this correspond to the probability distributions $p(s_1,\ldots, s_N|X_1,\ldots,X_N)$.

Now we introduce the correlation to the game. Mermin's non-locality scenarios make it possible to express non-locality in CQM. He defines a \emph{local
hidden variable} (LHV) model for a $n$-party state which consists of:\begin{itemize} \item a family of hidden states $|\lambda \rangle$, each of which
assigns for any measurement on each subsystems a definite outcome,

\item and , a probability distribution on these hidden states,
\end{itemize} which simulates the probabilities of that theory. As it is done for non-locality, the quantum correlation in a game can be formalized using the following
Born vector:

\begin{center}
\includegraphics[scale=0.7]{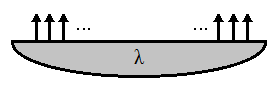}
\end{center} This Born vector represents the probability distribution on possible strategies. One can sample out $|\lambda )$ independently for each
strategy to obtain the game with classical correlation:

\begin{center}
\includegraphics[scale=0.7]{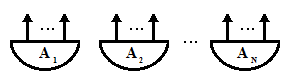}
\end{center} In \cite{Co3}, the authors showed these two possibilities are inequivalent by applying a function that computes the parity to both situations.
Hence, we can conclude that the games with quantum correlation and classical correlation are inequivalent in this case as they yield different probability
distributions.

\section{Conclusion}
Quantum technologies can be applied to a game either as randomizing devices or as a communication devices. The main aim of this paper is to introduce a
topological diagrammatics to the both types of quantum games. The mathematical foundation behind this approach is category theory and it is inspired by
categorical quantum theory (CQM) program which is originally founded to address the questions from quantum information and computation. The methods of CQM
usually makes computations and proofs easier \cite{Co3} and enables new generalizations like in the case of algorithms \cite{V}. This new presentation of
quantum communication approach, which is in the first part of this paper, might also help us to address certain questions:
\begin{itemize}
\item How does the strategy space of quantum moves affect the outcome of the game?

\item Does the initial state $|00\rangle$ and unitary operator $U$ have an important role?

\item Are there any other equivalent presentations of the game in CQM?

\item Can one represent different type of games in this formalism?
\end{itemize}

In the second part of this work, in order to represent the correlation in a quantum randomization scenario we used the result of Coecke and et. al.
\cite{Co3} on non-locality in CQM. In this case, we are mainly inspired by the work of Brenner and Linden \cite{BL} who pointed out the connection between
Bayesian Games and non-locality. They observed that the players who have access to quantum advice outperform the others. Our approach yields same result
via CQM stating that a game with quantum correlation is not equivalent to a game with classical correlation.


\begin{thebibliography}{99}
\bibitem{A} R. J. Aumann, \emph{Subjectivity and Correlation in Randomized Strategies}, J. Math. Econ \textbf{1} (1974).

\bibitem{AC} S. Abramsky and B. Coecke, \emph{A categorical semantics of quantum protocols},
Proceedings of the 19th Annual IEEE Symposium on Logic in Computer Science, pages 415-425, 2004. IEEE Computer Science Press.

\bibitem{AC2} S. Abramsky and B. Coecke, \emph{Handbook of Quantum Logic and Quantum Structures}, volume 2, chapter Categorical Quantum Mechanics.
Elsevier, 2008.

\bibitem{BL} N. Brunner and N. Linden, \emph{Bell nonlocality and Bayesian game theory}, Nat. Commun. \textbf{4}(2013)

\bibitem{CIA} J. M. Chappell, A. Iqbal, D. Abbott, \emph{$N$-player quantum games in EPR setting}, PLoS ONE7 \textbf{5}, art. no. e36404.

\bibitem{CI} T. Cheon, A. Iqbal, \emph{}, J. Phys. Japan \textbf{77}(2008), 024801.

\bibitem{Co} B. Coecke, \emph{The logic of entanglement: An
invitation}, Technical Report, University of Oxford, 2003. Computing Labaratory Research Report PRG-RR-03-12.

\bibitem{Co2} B. Coecke and R. Duncan, \emph{Interacting quantum observables: Categorical Algebra and diagrammatics},
New Journal of Physics. \textbf{13}(2011).

\bibitem{Co3} B. Coecke, R. Duncan, A. Kissinger, Q. Wang, \emph{Strong Complementarity and Non-locality in Categorical Quantum Mechanics}, Proceedings of
the 2012 27th Annual ACM/IEEE Symposium on Logic in Computer Science, LICS 2012, art. no. 6280443, pp. 245-254.

\bibitem{CK} B. Coecke, A. Kissinger, \emph{The compositional structure of multipartite quantum entanglement}, Lecture Notes in Computer Science 6199 LNCS,
pp. 297-308.

\bibitem{EWL} J. Eisert, M. Wilkens and M. Lewenstein, \emph{Quantum Games and Quantum Strategies}, Physics Review Letters, \textbf{83}(1999), 3077.

\bibitem{Gr} O. Grabbe, \emph{Introduction to Quantum Game Theory},

\bibitem{G} L. Grover, \emph{Quantum mechanics helps in searching for a needle in a haystack}, Physical Review Letters, \textbf{79}(1997), no. 2, 325-328.

\bibitem{KL} G. M. Kelly and M. L. Laplaza, \emph{Coherence for compact closed categories}, Journal of Pure and Applied Algebra, \textbf{19}(1980),
193-213.

\bibitem{La} S. E. Lansburg, \emph{Quantum Game Theory}, The Wiley Encylopedia of Operation Research and Management Science, 2011.

\bibitem{L} J. Lambek, \emph{From lambda calculus to Cartesian closed categories}, Academic Press, 376-402,1980.

\bibitem{M} D. Meyer, \emph{Quantum Strategies}, Physics Review Letters, \textbf{82} (1999), 1052-1055.

\bibitem{Me} N. D. Mermin, \emph{Quantum mysteries revisited,} American Journal of Physics \textbf{58}(1990), 731-734.

\bibitem{S} P. Sellinger, \emph{New Structures for Physics, chapter A Survey of Graphical Languages for Monodial Categories}, 289-355, Number 813 in
Lecture Notes in Physics. Springer, 2011.

\bibitem{V} J. Vicary, \emph{The Topology of Quantum Algorithms},
Proceedings-Symposium on Logic and Computer Science, art. no. 6571540, pp. 93-102.

\bibitem{vN} J. von Neumann, \emph{Mathematische Grundlagen der quantenmechanik}, Springer-Verlag, 1935.

\bibitem{VM} J. von Neumann and O. Morgenstern, \emph{Theory of Games and Economic Behaviour}, Princeton University Press, Second Ed., 1947.
\end{thebibliography}
\end{document}